\documentstyle{article}
\begin{document}
\large
\baselineskip=24pt
\title{{\bf HYDRODYNAMIC SPINODAL DECOMPOSITION:
\\GROWTH KINETICS
\\AND
\\SCALING FUNCTIONS}}
\author{F. J. Alexander, S. Chen, {\small AND} D. W. Grunau\\
  [.5cm]{\em Center for Nonlinear Studies}\\
	{\em and}\\
	{\em Theoretical Division}\\
	{\em and}\\
	{\em Earth and Environmental Sciences}\\
	{\em Los Alamos National Laboratory}\\
	{\em Los Alamos, NM 87545}\\[.4cm]}
\date{}
\maketitle

\vspace{0.4in}
\begin{center}
{\bf ABSTRACT}
\end{center}

We examine the effects of hydrodynamics on the late stage kinetics
in spinodal decomposition.  From computer simulations of a lattice
Boltzmann scheme we observe, for critical quenches, that single phase
domains grow asymptotically like $t^{\alpha}$, with $\alpha \approx .66$
in two dimensions and $\alpha \approx 1.0$ in three dimensions, both in
excellent agreement with theoretical predictions.

\pagebreak
\def\s{\sum_i}
\def\s{\sum_j}
\def\dt{\partial_t}
\def\x{{\bf x}}
\def\u{{\bf u}}
\def\f{{\bf f}}
\def\G{{\bf G}}
\def\grad{\nabla}
\def\sj{\sum_j}
\def\si{\sum_i}
\def\eps{\epsilon}
\newcommand{\e}[1]{{\bf e}_#1}

The understanding of phase segregation kinetics has been enhanced
over the last two decades by
the development of dynamic renormalization group methods,
an increase in computer resources,
and a host of experimental results with which to compare~\cite{GM,Furukawa}.
Despite this progress, there remain many open questions,
especially regarding the growth of single phase domains and the
scaling properties of the correlation or structure functions.
Since phase segregation, either spinodal decomposition or nucleation,
is a highly nonlinear process,  theories rely mostly on approximation
schemes, and exact results have been obtained only in certain limiting
cases~\cite{exact}.  Simulations designed to address these issues
typically require extensive computational effort and have not, in our
opinion,  provided conclusive answers to many of the most fundamental
questions.

Phase segregating systems fall into two classes: those with hydrodynamic
interactions (fluids) and those without (binary alloys, glasses).  The latter
class, has received far more attention both by computer simulation and by
theory\cite{GM,Furukawa}.  Only recently have there been attempts to carry out
computer simulations of phase segregating systems with hydrodynamic
interactions.  These include molecular dynamics (MD) simulations~\cite{MD},
direct numerical simulation of time-dependent Ginzburg-Landau (TDGL)
equations~\cite{FV}, cell-dynamical systems (CDS)~\cite{Shinozaki,Koga} and
lattice gas (LG)~\cite{RK,lglb} and lattice Boltzmann (LB)
models~\cite{Guns}.

While MD simulations accurately represent the dynamics
of real fluids, they are computationally demanding and at
present may not be able to access the late stage (scaling) regime of
spinodal decomposition.  The TDGL approach eliminates the exact Newtonian
particle dynamics in favor of a stochastic evolution governed by a
coarse-grained phenomenological free energy functional. One then solves
the concomitant system of Langevin equations which couple order parameter
fluctuations to hydrodynamic currents.  This approach requires extensive
numerical computation and sacrifices both the Navier-Stokes behavior
of the fluid and the interface phase dynamics.  CDS methods further
abstract the phase
segregation process by replacing the Langevin equations with a much less
numerically intensive dynamical map for the order parameter.
The hydrodynamics they incorporate, however, have either been approximated
by an Oseen tensor or by a coupling of the velocity field to pressure
and order parameter fluctuations.

LG and LB models, on the other hand, provide alternative computational
environments with which to study hydrodynamic phase segregation phenomena
without the introduction of {\em ad hoc} relations between the order
parameter fluctuations and the fluid dynamics.  What makes these schemes
so appealing is the natural way in which they can simulate the fluid
properties, the phase segregation and the interface dynamics simultaneously.
One drawback in the model presented here is that fluctuations only appear
in the initial conditions.  To make LB models more realistic tools with
which to analyze phase transition kinetics probably involves the
introduction of some type of thermal noise.

Several groups~\cite{RK,lglb,Guns}
have used LG and LB models to simulate
phase segregation phenomena, but their work focused primarily on
the {\em qualitative} features and did not address
the details of dynamical scaling of the structure function~\cite{Ernst}.
In this paper we focus on domain growth kinetics and scaling
properties of the phase segregation process. We present the results of
large-scale numerical simulations of one such immiscible LB model and
make quantitative comparisons with theory.

The LB method is a discrete, in space and time,
microscopic, kinetic equation description for the evolution of the
particle distribution function of a fluid~\cite{LB}.
The scheme described here is a modified version of
the immiscible fluid model introduced by Gunstensen {\it et al}~\cite{Guns}.
In this model the fluid has two components represented, for example,
by the colors red and blue. The microscopic dynamics of the
particle distribution function consists of four steps:
(1) free streaming, (2) collision,  (3) interface perturbation and
(4) recoloring.

In free streaming the fluid (both red and blue components)
moves to neighboring sites along the links of the
underlying lattice.  During the collision step these
densities then relax toward a local equilibrium state. In LB
schemes one is free to specify the local equilibrium state, and
the particular choice for this state is one which leads to
the Navier-Stokes equations in the long wavelength and low frequency
limit.

Let $f_i({\bf x}, t)$, $f_i^{r}({\bf x}, t)$ and $f_i^{b}({\bf x}, t) $
be the distribution functions for the total fluid, red (r)
fluid and blue (b) fluid, respectively,  at site ${\bf x}$ and time
$t$ moving along link in the $i$ direction.
Here $f_i({\bf x}, t) = f_i^{r}({\bf x}, t)
+ f_i^{b}({\bf x}, t)$,
where $i =  0, 1, \cdot\cdot\cdot,N$, and where $N$
is the number of velocity states at each site.
The $i = 0$ state represents the portion of the fluid at rest.
The kinetic or ``lattice Boltzmann'' equation for $f_i$ is written
\begin{eqnarray}
f_{i} ({\bf x}+ {\bf e}_{ i }, t+1) -
f_{i} ({\bf x},t) = \Omega^{c}_i({\bf x},t) +  \Omega^{p}_i({\bf x},t),
\end{eqnarray}
where $\Omega^{c}_{i}$ is the term representing the
rate of change of $f_{ i}$ due to collisions, and $\Omega^{p}_{i}$ is the
term representing the color perturbation. The vectors ${\bf e}_i$ are the
velocity vectors along the links of the lattice.

In this paper we use a triangular lattice (N = 6)  for two-dimensional
simulations with ${\bf e}_ {i}= (\cos{(2\pi (i-1)/6)},
\sin{(2\pi (i-1)/6)})$,
and a body-centered-cubic lattice (N = 14)
in three dimensions with ${\bf e}_ {i} \in ({\pm 1},0,0),
(0,{\pm 1},0)$, $(0,0,{\pm 1})$ and
$({\pm 1}, {\pm 1}, {\pm 1})$.
For computational efficiency, we have used the single time relaxation
model~\cite{BGK} with the linear collision operator:
\begin{eqnarray}
\Omega^{c}_{i} =  - \frac{1}{\tau}(f_i-f_i^{(eq)}).
\end{eqnarray}
where $\tau$ is the characteristic relaxation time, and
$f_i^{(eq)}$ is the local
equilibrium distribution
given in two dimensions by
\begin{eqnarray}
f_{0}^{(eq)} = \frac{\rho}{7} - \rho{\bf u}^2,
\end{eqnarray}
and
\begin{eqnarray}
f_{i}^{\rm (eq)} = \frac{\rho}{7}  + \frac{\rho}{3} {\bf e}_{ i}\cdot{\bf u}
+\frac{2\rho}{3}({\bf e}_{i} \cdot {\bf u})^{2} - \frac{\rho}{6} {\bf u}^{2},
\end{eqnarray}
and in three dimensions by
\begin{eqnarray}
f^{\rm (eq)}_{0} = \frac{\rho}{8}- \frac{\rho}{3}{\bf u}^2,
\end{eqnarray}
\begin{eqnarray}
f^{{\rm (eq)}}_{i} = \frac{\rho}{8}
+ \frac{\rho}{3}({\bf e}_{ i} \cdot {\bf u})
+\frac{\rho}{2}({\bf e}_{i} \cdot {\bf u})^{2} -  \frac{\rho}{6} {\bf u}^{2},
\end{eqnarray}
for ${\bf e}_{i}$ along the lattice axes, and
\begin{eqnarray}
f^{{\rm (eq)}}_{i} = \frac{\rho}{64}
+ \frac{\rho}{24}({\bf e}_{ i} \cdot {\bf u})
+\frac{\rho}{16}({\bf e}_{i} \cdot {\bf u})^{2} -  \frac{\rho}{48} {\bf u}^{2}
\end{eqnarray}
for ${\bf e}_{i}$ along the links to the corners of the cube.
In the above equations
$\rho({\bf x},t) = \sum_i f_i({\bf x},t)$ and $\rho{\bf u}({\bf x},t) = \sum_i
f_i({\bf x},t) {\bf e}_i$ are
the local density and momentum,  respectively.

The detailed forms
of the coefficients in Equations (3-7) are determined by the
conservation of mass and momentum, the constraints of Galilean invariance,
and a velocity independent, isotropic
pressure tensor.  It can be shown~\cite{cwds}
that the macroscopics of (1-4) and (1,2,5,6,7) correspond to the
incompressible Navier-Stokes equation in two and three dimensions,
respectively:
\begin{eqnarray}
\dt\u+(\u\cdot\nabla)\u={-1\over\rho}\nabla P+\nu\nabla^2\u\nonumber \\
\nabla\cdot\u=0,\nonumber
\end{eqnarray}
where $P$ is the pressure,
and $\nu$ is the kinematic viscosity. In two dimensions
$\nu=(2\tau -1)/8$, and in three dimensions $\nu = (2\tau -1)/6$.

Defining the local order parameter as $\psi({\bf x},t) = \sum_{i=0}^{N}
(f^r_i({\bf x},t)-f^b_i({\bf x},t))$, and the local color gradient $\G(\x)$ by
\begin{eqnarray}
\G(\x)=\sum_{i=1}^{N}\e{i}\left\{\psi(\x+\e{i})\right\},
\end{eqnarray}
we add the surface-tension inducing perturbation
\begin{eqnarray}
\Omega^p_i=A|\G|\cos 2(\theta_i-\theta_G)
\end{eqnarray}
to facilitate segregation and stabilize interfaces.
Here $\theta_i$ is the angle of lattice direction $i$, and
$\theta_G=\arctan(G_y / G_x)$ is the angle of
the local color gradient. The constant $A$ sets the
surface tension, $\sigma$, through $\sigma \sim A \tau \rho$.
Note that $\G$ is perpendicular to red-blue interfaces and its magnitude large
there, while in a homogeneous (color) region it approaches zero.
In the recoloring step we then maximize the color flux ${\bf H}=\sum_{i=1}^{N}
(f^r_i-f^b_i)\e{i}$
in the direction of the color gradient, $\G$,
by maximizing ${\bf H} \cdot \G$.
Recoloring conserves the individual color components and hence the total
density of the fluid.
It has been shown~\cite{Guns} that
 Laplace's law holds for this model.

For our simulations we carry out critical quenches with
$\sum_{{\bf x}} \psi({\bf x},t) = 0$.  The largest systems we simulated
were $(2048)^2$ in two
dimensions and $(128)^3$ in three dimensions.  Although we have investigated
the domain growth and scaling properties for a variety of lattice sizes and
parameters, we report on the domain growth and dynamical scaling
properties for only one set of parameters
in both two and three dimensions.  The results obtained with
smaller lattices and different parameters (surface tension and initial
fluctuations) are
consistent with the data presented here.

We initialize the lattice (in both two and three dimensions)
with $\langle \psi ({\bf x}) \rangle = 0$ and $\langle {\bf u} \rangle = 0$
with small local fluctuations, where
the angle brackets $\langle \rangle$ signify an average over the lattice.
The system then evolves according
to the dynamics outlined above. In two dimensions the surface tension
inducing parameter $A=.01$, and the average
fluid density per site is $\rho = 2.1$, and in three dimensions $A=.001$ and
$\rho = 2.4$. As the system evolves, single color domains form and grow while
the total fluid undergoes a Navier-Stokes evolution.   In Figure 1 we
show a ``snapshot'' (for a system of size $(256)^2$)
of a two dimensional system at time $t=1000$.

One convenient way to characterize the growth kinetics during
the segregation process is through the order parameter correlation function,
$G({\bf r},t) = \langle \psi ({\bf r})\psi (0) \rangle -
\langle \psi \rangle ^2$, averaged over shells of radius $r$.
One can then define the domain size, $R(t)$, as the first zero of the function
$G (r,t)$.  The Fourier transform of $G$ is then
the structure factor $S(k,t)$.  As time evolves the structure factor becomes
more sharply peaked, and
its maximum moves to smaller values of the wavenumber, $k$.
In a wide variety of phase segregating systems $S(k,t)$
has been observed to follow the dynamic scaling relation at late
times~\cite{GM}:
\begin{eqnarray}
S(k,t) \approx R^d (t) F(x),
\end{eqnarray}
where $x= kR(t)$, $d$ is the spatial dimension and $F(x)$
is the structure factor.

In Figure 2(a) we plot $R(t)$ for the two dimensional model.
The data indicate that $R(t) \sim t^{.66}$.  This exponent is in excellent
agreement with the generally accepted theoretical prediction of
$t^{2/3}$~\cite{Furukawa} and the numerical simulations of Ferrel and
Valls~\cite{FV} which find $R(t) \sim t^{\alpha}$ where
$\alpha \approx .65$ and $\alpha \approx .69$ for
systems with and without thermal modes respectively.

In Figure 3 we show the scaled structure factor, $F(x)$, for our
two dimensional simulations.  For $x > 2$ we find that $F(x) \sim x^{-3.3}$
in reasonable
agreement with Porod's law which predicts that $F(x) \sim x^{-(d+1)}$
in this region.  However, for $x <1$, we observe that $F(x) \sim x^{2.2}$.
This is in sharp contrast with the theoretical arguments of
Furukawa~\cite{Furukawa}
and Yeung~\cite{Yeung} which predict that $F(x) \sim x^{4}$.  Our findings for
$x \ll 1$ appear to corroborate other recent numerical simulations
of hydrodynamic spinodal decomposition in two dimensions.  Shinozaki and
Oono in their CDS model observe that $F(x) \sim x^2$.
They conjecture that this might be a result of finite size effects coupled
to fluctuations of domain walls due to long range hydrodynamic interactions.

In Figure 2(b) we plot $R(t)$ for our three dimensional model.
Here we find that the domain growth is approximately linear in time
with $R(t) \sim t^{1.0}$. This is in excellent agreement with the
theoretical predictions of Furukawa~\cite{Furukawa} and
Siggia~\cite{Siggia} and with the (TDGL) numerical
simulations of Ferrel and Valls and the CDS simulations of Koga and Kawasaki.
It disagrees, however,  with recent MD simulations of Ma et. al~\cite{MD}
who observe
a domain growth $R(t) \sim t^{.55 \pm .05}$. This latter exponent
may not reflect the growth in asymptotic time regime.

In Figure 4. we plot $F(x)$ for the three dimensional model.
As in two dimensions we find good agreement with Porod's law
with $F(x) \sim x^{-4.3}$ for $x > 3$.
Again, though we find marked disagreement
with the small $x$ predictions of $F(x)$ which call for
$F(x) \sim x^{4}$.  In particular
we find that $F(x) \sim x^{2.3}$.
This result is consistent with recent light scattering data from spinodal
decomposition of isobutyric acid and water~\cite{Kubota} which
also indicates a reduced exponent $(\approx 2)$, but differs from
the results of Bates and Wiltzius~\cite{Wiltzius}.

We summarize by pointing out that the results of our simulations tend to
be consistent with existing theories of domain growth in both two
and three dimensions.  There is also good agreement with Porod's law
in both of these cases.  However, we find a marked discrepancy with
the theoretical predictions of Yeung~\cite{Yeung} and Furukawa~\cite{Furukawa}
for the small $x$ behavior of the scaled structure factor.
Moreover, there seems to be a consistent
deviation from these theories among some recent numerical~\cite{Shinozaki}
and some
experimental work~\cite{Kubota}.   In light of these results, we believe
that the question of the small $x$ behavior of the scaled structure factor
is still open and that it is necessary to develop a more complete theory
of dynamical scaling which includes hydrodynamic interactions.

The current model simulates fluids for deep (near zero temperature) quenches.
As a result, this precludes an analysis of binary fluids near their critical
points. To study critical dynamics of binary fluids
requires some notion of temperature.
An extension of the model described here might include
a stochastic term which mimics the effects of
thermal noise (Landau-Lifshitz fluctuating hydrodynamics)~\cite{Ladd}
or a kinetic temperature as in Reference~\cite{ACS} with
which
one can control the segregation process by a local temperature.
Without such a noise term it is unlikely that this model can simulate
off critical quenches.

The model in this paper is ideally suited for the simulation
of hydrodynamic phase segregation in high Reynolds number
flows~\cite{cwds}.   Moreover, the LB scheme can easily
simulate spinodal decomposition in
systems with complicated boundaries,
stirring, and the effects of wettability.

We thank K. L. Diemer, G. D. Doolen, K. G. Eggert, G. L. Eyink, S. Habib,
S. A. Janowsky, J. L. Lebowitz, R. Mainieri, and J. D. Sterling for
useful discussions.  We are also indebted to the referee for helpful
suggestions and comments.  Eugene Loh was involved in the original
CM-2 code development. This work was supported by the U. S. Department of
Energy at Los Alamos National Laboratory.
Numerical simulations were carried out using the computational resources at
the
Advanced Computing Laboratory at the Los Alamos National Laboratory.


\begin{center}
\underline{{\bf FIGURE CAPTIONS}}
\end{center}

Figure 1: Typical configuration for the two-dimensional model at
time $t=1000$.  The system size is $(256)^2$, and the other
parameters are as in the text.
In both (a) and (b)
The black region represents sites with positive
order parameter.

Figure 2: Time dependence of average domain size
for (a) two-dimensional and (b) three-dimensional systems.
Both two and three dimensional data is averaged over
2 independent runs.

Figure 3: Scaled structure function $F(x)$ for 2 dimensions at various times:
$t=1000$ ($\triangle$);
$t=2000$ ($\Box$), and
$t=3000$ ($\Diamond$). The data is averaged over
2 independent runs.

Figure 4: Scaled structure function $F(x)$ for 3 dimensions at various times.
$t=1200$ ($\triangle$);
$t=2400$ ($\Box$), and
$t=3200$ ($\Diamond$). The data is averaged over
2 independent runs.

\end{document}